\documentclass[10pt, a4paper,twocolumn]{article}
\usepackage{xurl}
\usepackage[square,numbers,sort&compress]{natbib}
\usepackage{abstract}
\usepackage{lipsum}
\usepackage{amsmath}
\usepackage{graphicx}
\usepackage{tabularx}
\usepackage{soul}
\usepackage{geometry}
\usepackage[normalem]{ulem}
\usepackage[dvipsnames]{xcolor}
\usepackage{authblk}
\usepackage{array}

\usepackage{hyperref}
\hypersetup{colorlinks=true, urlcolor=blue, linkcolor=blue, citecolor=blue}

\title{Neutral and charged pion Form Factors in the intermediate-energy region from double-dilaton HQCD model}
\author[]{H\'ector Cancio}
\author[]{Pere Masjuan}
\affil[]{\textit{\small{Grup de F\'isica Te\`orica, Departament de F\'isica, Universitat Aut\`onoma de Barcelona, and Institut de F\'isica d’Altes Energies (IFAE), and The Barcelona Institute of Science and Technology (BIST), Campus UAB, 08193 Bellaterra (Barcelona), Spain}}}

\begin{document}
\twocolumn[
\begin{@twocolumnfalse}
\maketitle
\begin{abstract}
      We compute the Form Factors of both neutral and charged pion using a non-perturbative running of the strong-coupling constant $\alpha_s$ obtained using a double-dilaton Holographic QCD model. These form factors remain poorly understood in the intermediate-energy region, which marks the transition between low- and high-energy physics. In particular, experimental data for the neutral pion Form Factor exhibits a deviation from the expected asymptotic behavior, and the charged pion form factor remains comparatively less explored. To address these issues, we employ the pion distribution amplitude formalism to investigate the Form Factor behavior in this intermediate regime. Our results suggests that non-perturbative physics of the strong interaction is relevant even at energy scales traditionally considered perturbative, implying that the perturbative regime could occur at higher energies than previously thought. Finally, our approach allows us to study isospin-breaking effects through the quadratic pion mass difference.
\end{abstract}
\end{@twocolumnfalse}
]

\vspace{5em}

\section{Introduction}
The neutral and charged pion form factors in the space-like momentum domain have been key objects for studying the inner structure of the lightest hadron, the pion, and also serve as a probe to study non-perturbative aspects of QCD. At this moment, there are different sources of experimental data to determine whether the QCD perturbative predictions are satisfied in Nature for the neutral (\cite{BaBar,Belle,CELLO,CLEO}) and the charged (\cite{NA7,JLAB1,JLAB2,Dally1,Dally2,Dally3,Bebek74,Bebek76,Bebek78}) pion form factors.

There is interest in the intermediate-energy region and in the transition between IR and UV physics, and the role of the hadronic scale $\Lambda_{QCD}$. Remarkably is BABAR \cite{BaBar} and BELLE \cite{Belle} Collaborations' data for the neutral pion transition form factor, in which a non-negligible statistical deviation with the perturbative QCD asymptotic limit \cite{LepageBrodsky,Radyushkin,EfremovRadyushkin,EfremovRadyushkin2,BrodskyFarrar,BrodskyFarrar2,FarrarJackson}, can be observed. In terms of the charged-pion Form Factor, the experimental data also show a deviation from its perturbative asymptotic decay \cite{LepageBrodsky}. Various approaches have been used to study both form factors and pion decays in general. These include numerical calculations \cite{Masjuan:2015cjl,GerardinMeyer,RayaBashir}, decays \cite{Masjuan:2015lca, EscribanoDalitz}, perturbative QCD analyses \cite{ShenPower}, and light-front QCD \cite{ChoiLF,StefanisLC,SernaLF}.

In this letter, we want to address whether the form factors at intermediate regions, well above $\Lambda_{QCD}$, can be described by non-perturbative physics described through a non-perturbative strong running coupling.
To this end, we employ the results from \cite{CancioMasjuan} to have a well-defined strong coupling at all energies and study its effects at these intermediate energy scales. As such, we use the pion distribution amplitude formalism to implement our method. This consists of using a high-energy expansion of the pion distribution amplitude, which translates into an expansion for the form factors that depend on the strong coupling, and then fitting the experimental data at low energies. That fit is performed with the help of a Padé approximant after finding the appropriate matching conditions between the two regimes \cite{NogueraVento}. Our objective is to see if the non-perturbative strong coupling normalized to data at low energies has any effect at intermediate energy scales when requesting matching with the very-high energy region.
\\
\\
This letter is organized as follows. In Section 2, we present the pion distribution amplitude formalism and the model we will use for the strong coupling constant $\alpha_s$ defined at all energies. In Section 3, we present the main results of our work for the neutral and charged pion form factors, and finally, in Section 4, we conclude with the implications of our findings.
\section{Pion Form Factors and Pion Distribution Amplitudes} 
We employ the distribution amplitudes (DAs) $\phi_{\pi^0}$ and $\phi_{\pi^{\pm}}$ to define the neutral $F_0(Q^2)$ and charged $F_{\pi}(Q^2)$ pion form factors, respectively, as: 
\begin{equation}
\label{eqn:neutralFFdef}
Q^2F_0(Q^2)=\frac{\sqrt{2}f_{\pi}}{3}\int_0^1 \frac{dx}{x}\phi_{\pi^0}(x,Q^2),
\end{equation}
\begin{equation}
\label{eqn:chargedFFdef}
Q^2F_{\pi}(Q^2)=\frac{8\pi f_{\pi}^2 \hat{\alpha}_s(Q^2)}{9}\left(\int_0^1\frac{dx}{x}\phi_{\pi^{\pm}}(x,Q^2)\right)^2\, .
\end{equation}
Here, $f_{\pi}=0.131 \text{ GeV}$ is the pion decay constant and $\hat{\alpha}_s(Q^2)$ is the running strong coupling constant. 
We assume an expansion for high $Q^2$ in terms of Gegenbauer polynomials $C^{3/2}_{2n}(2x-1)$ inspired by the one derived in \cite{LepageBrodsky}, for neutral and charged pion form factors, respectively: 
\begin{equation}
\label{eqn:NeutralPDAExpansion}
\phi_{\pi^0}(x,Q^2)=\phi_{\text{asym}}(x)\sum_{n=0}^{\infty} b_{n}\hat{\alpha}_s(Q^2)^{\gamma_n}C^{3/2}_{2n}(2x-1),
\end{equation}
\begin{equation}
\label{eqn:ChargedPDAExpansion}
\phi_{\pi^\pm}(x,Q^2)=\phi_{\text{asym}}(x)\sum_{n=0}^{\infty} \tilde{b}_{n}\hat{\alpha}_s(Q^2)^{\gamma_n}C^{3/2}_{2n}(2x-1),
\end{equation}
where $x$ is the momentum fraction and $\phi_{\text{asym}}(x)=6x(1-x)$ is the asymptotic limit of the functions $\phi_{\pi^0}(x,Q^2)$ and $\phi_{\pi^\pm}(x,Q^2)$ when $Q^2\to\infty$. $b_n$ and $\tilde{b}_{n}$ are the expansion coefficients and $\gamma_n$ are anomalous dimensions. We normalize the two pion DAs such that:
\begin{equation}\label{normalizedDAs}
\int_0^1 dx \, \phi_{\pi^0}(x,Q^2)=\int_0^1 dx \, \phi_{\pi^\pm}(x,Q^2)=1.
\end{equation}

In this work, we use for $\hat{\alpha}_s$ the expression obtained in our previous work \cite{CancioMasjuan}.  It has important differences with respect to the standard perturbative $\alpha_s$: Our $\hat{\alpha}_s$ has an infrared fixed point and a smooth matching with the pQCD calculation at $Q_0=3.79 \text{ GeV}$. It embraces a description at all energies. Since the matching point is located above the usual $\Lambda_{QCD}\sim 1 \text{ GeV}$ scale, we understand that we carry on the non-perturbative physics above its domain, and thus, it influences the intermediate-energy region of both pion form factors. Since this is the region believed to be already in the pQCD landscape, the observed deviation may then be studied as an inference of the non-perturbative physics in the perturbative one by using such $\hat{\alpha}_s$ prescription.  We understand that this strong coupling constant could also be used to clarify the type of decay of the charged pion form factor at high $Q^2$ to its asymptotic pQCD result. Therefore, the main objective of this letter is to test the implications of a strong coupling constant defined at all energies in the form factors, in particular in the intermediate-energy region.

The exact expression for the model we use for the strong coupling constant is:
\begin{equation}
\label{eqn:DDSW}
\hat{\alpha}_s(Q^2):=\frac{\alpha_s(Q^2)}{\pi}=\frac{c}{1-a \text{ sech}^{4/5}(b(Q^2+1))}
\end{equation}
where the constants were determined as $a=0.962(1)$, $b=0.297(7) \text{ GeV}^{-2}$ and $c=0.070$ (see details in \cite{CancioMasjuan}).
The neutral and charged pion form factors have the following asymptotic limits, respectively:
\begin{equation}
\label{eqn:asympLim}
\lim_{Q^2\to\infty}Q^2F_0(Q^2)=\sqrt{2}f_{\pi},
\end{equation}
\begin{equation}
\label{eqn:chargedAsympLim}
\lim_{Q^2\to\infty}Q^2F_{\pi}(Q^2)=8\pi f_{\pi}^2\alpha_s(Q^2).
\end{equation}
Inserting the expansion at high $Q^2$ given by Eq.(\ref{eqn:NeutralPDAExpansion}) and Eq.(\ref{eqn:ChargedPDAExpansion}) into the definition Eq.(\ref{eqn:neutralFFdef}) and Eq.(\ref{eqn:chargedFFdef}), we end up with the following parametrizations for the neutral and charged pion transition form factors, respectively,
\begin{equation}
\label{eqn:neutralFFExpansion}
Q^2F_0(Q^2)=\frac{\sqrt{2}f_{\pi}}{3}\sum_{n=0}^{\infty}c_n\hat{\alpha}_s(Q^2)^{\gamma_n},
\end{equation}
\begin{equation}
\label{eqn:chargedFFExpansion}
Q^2F_\pi(Q^2)=8\pi f_{\pi}^2\hat{\alpha_s}(Q^2)\left|\sum_{n=0}^{\infty}\tilde{c}_n\hat{\alpha}_s(Q^2)^{\gamma_n}\right|^2.
\end{equation}

The $c_n$ and $\tilde{c}_n$ coefficients in Eqs.(\ref{eqn:neutralFFExpansion}), (\ref{eqn:chargedFFExpansion}) are related to $b_n$ and $\tilde{b}_n$ from Eqs.(\ref{eqn:NeutralPDAExpansion}), (\ref{eqn:ChargedPDAExpansion}) by  $b_n=c_n/3$ and $\tilde{b}_n=\tilde{c}_n$, respectively.  Eq.(\ref{eqn:asympLim}) imposes $c_0=3$. For the charged form factor, the asymptotic limit Eq.(\ref{eqn:chargedAsympLim}) does not fix the first coefficient $\tilde{c}_0$ in Eq.(\ref{eqn:chargedFFExpansion}); therefore, we will need a fit to experimental data to obtain this coefficient. Since it is a high $Q^2$ expansion, we use only the experimental data from \cite{Bebek74}, \cite{Bebek76}, and \cite{Bebek78}. 

Data on the neutral pion form factor at intermediate energies seem not to fully agree with the pQCD calculation \cite{LepageBrodsky}. Does that mean the perturbative prediction is only valid at higher energies than previously thought? Another possibility would be that uncertainties in the experimental data points are underestimated, or that the theoretical asymptotic limit has an uncertainty and should be interpreted as a band \cite{MasjuanArriolaBroniowski}. Different approaches have been used to study this transition form factor. At low energies, Chiral Perturbation Theory gives corrections to the decay $\pi^0\to e^+e^-\gamma$ \cite{Kampf}, but experimental data does not extend to sufficiently low $Q^2$ \cite{CELLO}. On the other hand, the Dalitz decay of $\pi^0\to e^+e^-\gamma$ in the timelike region was studied originally in \cite{Dalitz}, while radiative corrections were studied in \cite{Joseph}. It provides useful information to study the neutral pion form factor since the interaction between photons and the pion is not point-like. Other approaches include dispersive approaches using the anomaly sum rule \cite{Khlebtsov, Melikhov, Klopot} and light-cone sum rules \cite{Stefanis}, \cite{Mikhailov} and \cite{ChoiLF,StefanisLC,SernaLF} as stated in the introduction.

In this letter, we propose to use Padé Approximants as in \cite{Masjuan}  to extend the description to low energies, thus covering the energy region for which data are available, and merge with the pQCD description coming from high energies. The procedure requires a matching that we describe in the following. A Padé Approximant (PA) $P_M^N(x)$ of order $[N,M]$ of a function $f(x)$ is a rational function as the quotient of two polynomials $R_N(x)$ and $Q_M(x)$ of order $N$ and $M$ respectively, such that its Taylor expansion at $x=0$
\begin{equation}
P_M^N(x)=\frac{R_N(x)}{Q_M(x)}=\frac{a_0+a_1 x+...+a_N x^N}{1+b_1x+...+b_M x^M}
\end{equation}
satisfy \cite{Baker}:
\begin{equation}
Q_M(x)f(x)-R_N(x)=\mathcal{O}(x^{N+M+1}).
\end{equation}

PAs are used in this work to fit and extrapolate the information from the data on the form factor up to an energy scale at which the pQCD result matches smoothly. To perform this matching, we introduce a higher-order term in the polynomial of the numerator of the PA to impose matching with the high-energy form factor. This means we will impose continuity and differentiability at some unknown point $Q_0^2$ to solve the system of equations.

\section{Results for \texorpdfstring{$F_0$}{F0} and \texorpdfstring{$F_{\pi}$}{Fpi} and their DAs}

\subsection{Pion Transition Form Factor \texorpdfstring{$F_0(Q^2)$}{F0(Q2)}} 

\begin{figure}
\includegraphics[scale=0.5]{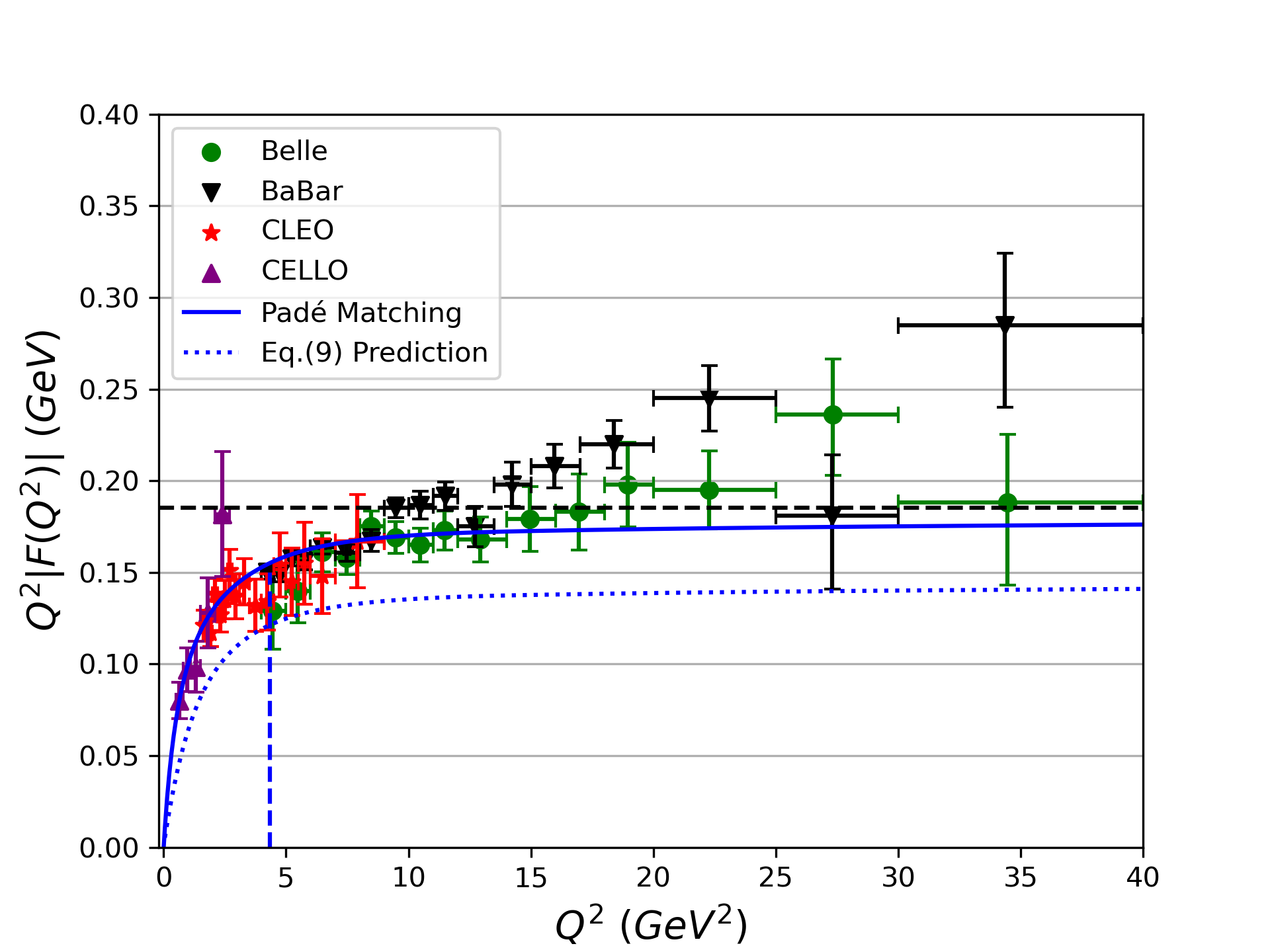}
\caption{Comparison between experimental data of $Q^2F_0(Q^2)$ from CELLO Col. \cite{CELLO}, purple triangles, CLEO Col. \cite{CLEO}, red stars, BABAR Col. \cite{BaBar}, black triangles, and BELLE Col. \cite{Belle}, green dots; and the model prediction either fitted to data using Eq.(\ref{hybrid}) (solid blue line) or without fitting to them using Eq.(\ref{eqn:neutralFFExpansion}) with $n=1$ (dotted blue line). Dashed blue vertical line indicates the matching $Q^2_0$ point.}
\label{fig:neutralFF}
\end{figure}

\begin{figure}
\includegraphics[scale=0.5]{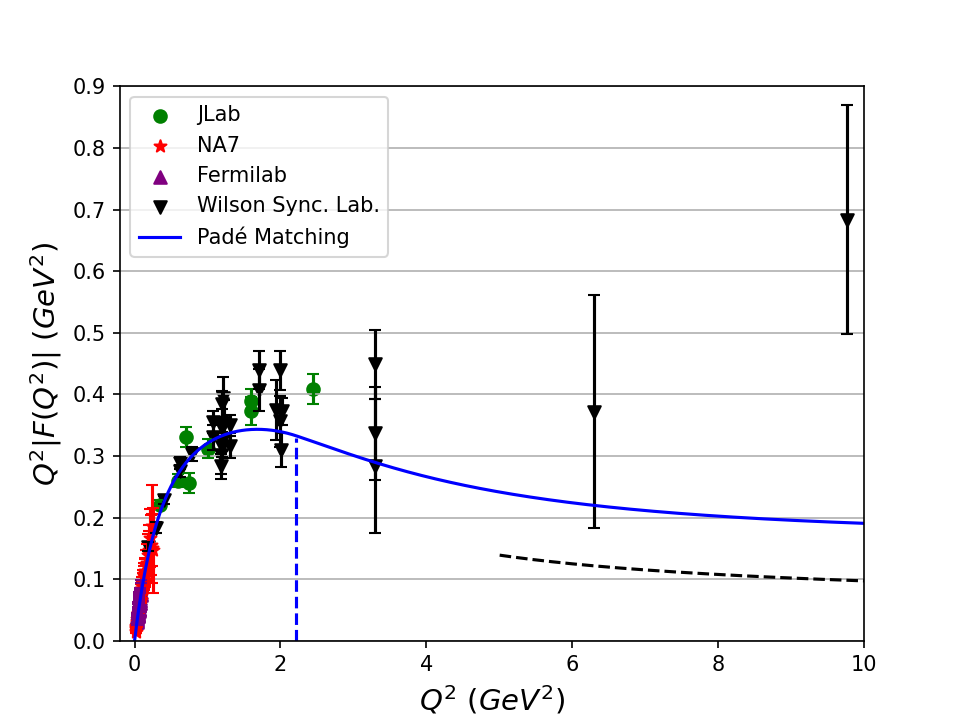}
\caption{Charged pion form factor. The solid blue curve is the result of using Eq.(\ref{eqn:chargedFFExpansion}) with $n=1$ at high energies with a matching procedure described in the text at low energies, see Sec.(\ref{subsec:3.2}). Experimental data in green dots (JLab \cite{JLAB1}, \cite{JLAB2}), red stars (NA7 \cite{NA7}), purple triangles (Fermilab \cite{Dally1}, \cite{Dally2}, \cite{Dally3}) and black triangles (Wilson Sync. Lab. \cite{Bebek74}, \cite{Bebek76} and \cite{Bebek78}). Dashed blue vertical line points the matching condition. Black dashed curve emulates the pQCD result.}
\label{fig:chargedFF}
\end{figure}

Fig.(\ref{fig:neutralFF}) compares the prediction for $Q^2F_0(Q^2)$ from Eq.(\ref{eqn:neutralFFExpansion}) at $n=1$ using our parametrization for $\hat{\alpha}_s$  with $c_0=3$ and $c_1=-c_0=-3$ versus experimental data from CELLO, CLEO, BABAR, and BELLE Collaborations, as a dotted-blue line. The value $c_0=3$ comes from imposing the asymptotic limit Eq.(\ref{eqn:asympLim}), and $c_0=-c_1$ imposing $Q^2F_0(Q^2)=0$ at $Q^2=0$. The condition $c_0=-c_1$, which is $c_1<0$, imposes that $Q^2F_0(Q^2)$ must be strictly increasing in $Q^2$, and then, cannot have a maximum before reaching the asymptotic limit Eq.(\ref{eqn:asympLim}). As a consequence, the asymptotic limit is reached "by" below and very far in $Q^2$, and not "by" above, as the data may seem to suggest. Our expansion in $\hat{\alpha}_s(Q^2)$ allows us to extend the region of validity of the perturbative expansion to lower energies, well within the region where data are available. This is a unique characteristic of our all-$Q^2$-defined $\hat{\alpha}_s(Q^2)$, which is not possible with similar expansions in the pQCD strong coupling constant. Yet, with this naive holographic-QCD prediction, fully theoretical, experimental data are not reproduced. 

We remark that this \textit{naive} description contains only two ingredients: an asymptotic form factor description convoluted with a strong coupling constant $\hat{\alpha}_s(Q^2)$ obtained in Ref.\cite{CancioMasjuan} with the unique ability to extend the large-$Q^2$ regime down to $Q^2\to 0$. Even though data cannot be reproduced, it is remarkable that the description covers the full $Q^2$ range with an up-to ${\cal O}(\alpha_s)$ expression. Including ${\cal O}(\alpha_s^2)$, i.e., a $c_2$ term in Eq.(\ref{eqn:neutralFFExpansion}), may improve the description. We, however, have no theoretical constraint to fix this coefficient univocally. $c_2$ term, being of ${\cal O}(\alpha_s^2)$, contributes very little to the description of the form factor.

An alternative upgraded model to data may let the parameters $c_0$ and $c_1$ free, without imposing the limit constraints and fitting all the experimental data available. To do that, and to ensure data are included in the description, we devise a hybrid model in which low energies are described via a Padé approximation $P^N_M(Q^2)$ fitted to data following Ref.\cite{Masjuan}, and high energies via the asymptotic form factor description. Then, a matching point -where continuity and derivability are imposed- must be found. Our hybrid model is defined as:
\begin{equation}\label{hybrid}
Q^2F_0(Q^2) =
\begin{cases}
P^N_M(Q^2)     & \text{if } Q^2 \leq Q_0^2, \\
\text{Eq.}(\ref{eqn:neutralFFExpansion})        & \text{if } Q^2 > Q_0^2,
\end{cases}
\end{equation}
\noindent
being $Q_0^2$ the unknown matching point. In this case, at low energies \cite{CELLO}, \cite{CLEO} we use a $P_1^2$ approximant of the form:
\begin{equation}
P_1^2(Q^2)=\frac{a_1Q^2+a_2Q^4}{1+b_1Q^2}.
\end{equation}
This Padé $P^2_1(Q^2)$ is used to obtain the best possible normalized $\chi^2$. $a_0=0$ is imposed to satisfy $Q^2F_0(Q^2)=0$ at $Q^2=0$. First, we fit the low-energy data with $a_2=0$, obtaining $a_1=0.224\pm0.037$ and $b_1=1.25\pm0.28$. For the high-energy regime, we obtain using Eq.(\ref{eqn:neutralFFExpansion}) with $n=1$, $c_0=c_0^{\rm fit}=3.616\pm 0.083$ and $c_1=c_1^{\rm fit}=-3.2\pm0.20$. 
Later, we go back to our $P^2_1(Q^2)$ and consider $a_2\neq 0$, and we determine it by imposing matching conditions at an unknown matching point $Q_0^2$ with our parametrization at high energies with an error of $10^{-11}$:
\begin{equation}
P_1^2(Q_0^2, a_2)=Q_0^2F_0(Q_0^2)
\end{equation}
\begin{equation}
\frac{d P_1^2(Q^2)}{dQ^2}\Bigg{|}_{Q=Q_0^2, a_2}=\frac{d(Q^2F_0(Q^2))}{dQ^2}\Bigg{|}_{Q=Q_0^2}
\end{equation}
By solving this system of equations numerically, we obtain $a_2=1.4\cdot 10^{-3}$ and a matching point at $Q_0^2=4.35\text{ GeV}^2$. Since the unknown $Q_0^2$ matching point splits what data points belong to low energies and what to high energies, we solve the system iteratively.  Notice that at this intermediate energy, only CLEO and CELLO Col. data are included in the PA fit, while the asymptotic description is fed from the rest.

Convergence of PAs to the function under study improve as $N,M \to \infty$. Even though we fit experimental data (and different PAs would yield similar results within the domain of experimental data), the extrapolations at high and low energies will depend on the order of the approximant 
(cf. \cite{Masjuan,Escribano:2015yup,Escribano:2015nra}). However, our exercise now is different as we match at a particular $Q^2_0$ with another description at high energies for which only 2 parameters are used. Then, left and right descriptions must be information-equivalent and, on top, a criterion to define $Q^2_0$ is still needed. 

We decide, then, to simplify the scenario by imposing best-$\chi^2$ fit together with an Akaike Information Criterion (AIC) where unnecessary model complexity with respect to data description is penalized. The best fit within the AIC selected the $P^2_1(Q^2)$ approximant as the least redundant, taken into account the obtained $Q^2_0$. Even though other approximants would be valid as well, and maybe with smaller theoretical uncertainty, the $P^2_1(Q^2)$ maximizes the available information.

The combined fit is achieved with $\tilde{\chi}^2=\chi^2/DOF = 2.7$ ($DOF$ as degrees of freedom). This minimized $ \tilde{\chi}^2$ is large, $>1$. This suprising result is understood after a close inspection to the individual normalized $\chi^2$ for each collaboration: $\tilde{\chi}^2_{\text{CLEO}} = 0.71$, $\tilde{\chi}^2_{\text{CELLO}} = 0.91$, $\tilde{\chi}^2_{\text{BELLE}} = 0.88$, and $\tilde{\chi}^2_{\text{BABAR}} = 5.14$. In terms of $\tilde{\chi}^2$, collaborations seem to be compatible only at a given level. We remark that the PA description can surpass the asymptotic limit at a finite energy if needed to improve consistency \cite{Masjuan}. The asymptotic description reaches, however, the asymptotic result by above. Even though the PA could overpass the asymptotic limit and then recover it by below to match the asymptotic description, such possibility is discarded by the AIC criterion. The method is flexible, but the minimized $\chi^2$ criterion and the AIC prefers reaching the asymptotic limit by below.

We finally remark that a fit without PA at low energies, yet possible, enhances $\tilde{\chi}^2$, in contrast to the case of the $F_{\pi}(Q^2)$, as we will see in the next subsection. A complete study of all other PAs was duly pursued, a report of which can simply be summarized by saying the second-best option $P_0^2$ has a matching point $Q_0^2=0.023\text{ GeV}^2$ outside the experimental data region, so it was discarded as a consequence of that. The rest yielded worse scenarios. Nevertheless, all in all, it reinforces the idea Eq.(\ref{eqn:neutralFFExpansion}) offers a parametrization that arrives at low energies.

\begin{figure}
\includegraphics[scale=0.5]{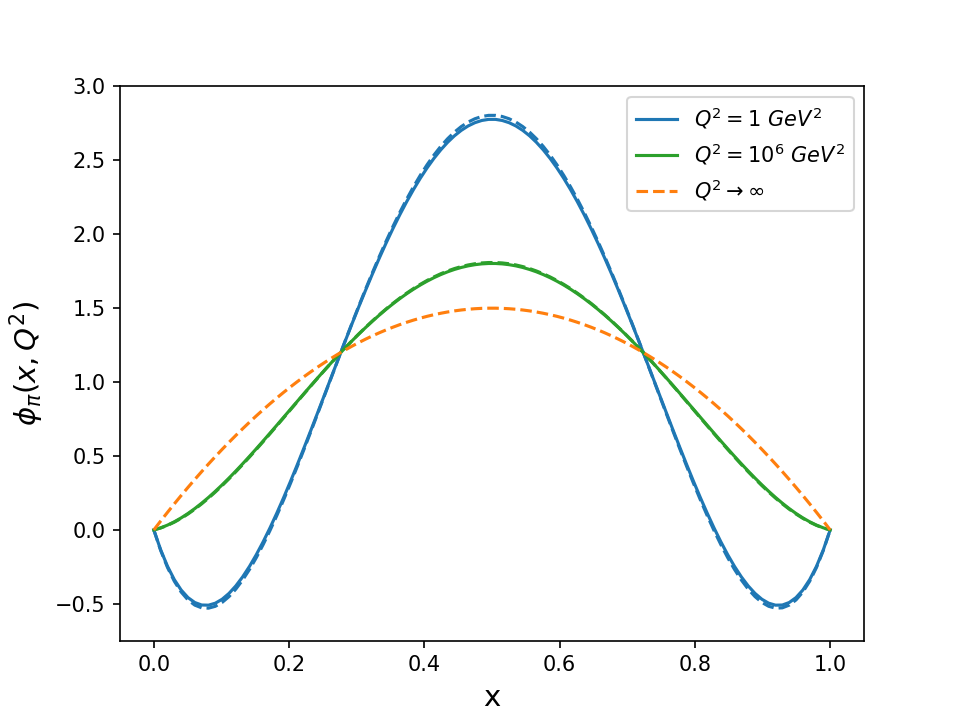}
\caption{Normalized neutral (charged) pion distribution amplitude in discontinuous (solid) blue at $Q^2=1 \text{ GeV}^2$, in green at $Q^2=10^6 \text{ GeV}^2$, and the asymptotic limit $Q^2\to \infty$ in orange.}
\label{fig:PDA}
\end{figure}

The previous solution with $c_1=-c_0=3$ is not compatible with the new one within uncertainties. This second solution is shown in Fig.(\ref{fig:neutralFF}) as a solid-blue line. 
A graphical account of the corresponding DA can be seen in Fig.(\ref{fig:PDA}), where dashed lines correspond to the scenario $c_0=c_0^{\rm fit}=3.616\pm 0.083$,  $c_1=c_1^{\rm fit}=-3.2\pm0.20$ and the solid line corresponds to the charged pion DA, explained in the following section.

Finally, the $\phi_{\pi^0}$ corresponding to $F_{0}$ is depicted in Fig.(\ref{fig:PDA}), properly normalized  following Eq.(\ref{normalizedDAs}).

\subsection{Electromagnetic Form Factor \texorpdfstring{$F_{\pi}(Q^2)$}{Fpi(Q2)}}
\label{subsec:3.2}
For $F_{\pi}(Q^2)$, the asymptotic limit Eq.(\ref{eqn:chargedAsympLim}) does not fix our expansion parameters, so we consider a fit to the experimental data with the expansion coefficients as parameters from Eq.(\ref{eqn:chargedFFExpansion}). For this task, we employ only the experimental data on the high-energy region, data from \cite{Bebek74}, \cite{Bebek76}, and \cite{Bebek78}. As a result, we obtain $c_0=3.075\pm0.055$ and $c_1=-2.664\pm0.074$. Then, inspired by Ref. \cite{Masjuan:2008fv}, we use again a hybrid model as in Eq.(\ref{hybrid}) for the low-energy regime. There we fit the experimental data from \cite{NA7}, \cite{JLAB1}, and \cite{JLAB2} with a $P^4_1(Q^2)$ to $F_{\pi}(Q^2)$ of the form:
\begin{equation}
\label{eqn:PadéApprox}
P_1^4(Q^2)=\frac{a_1Q^2+a_2Q^4+a_3Q^6+a_4Q^8}{1+b_1Q^2}
\end{equation}
We follow the same strategy as in the $F_0(Q^2)$ case. This precise $P_1^4(Q^2)$ is selected to obtain, after a matching procedure with the high-energy result, a $\tilde{\chi}^2$ function as close to $1$ as possible. Note that we impose $a_0=0$ as $Q^2F(Q^2)=0$ at $Q^2=0$. Then we fit the data imposing now $a_4=0$. This gives the parameters $a_1=1.004\pm0.012$, $a_2=-0.18\pm 0.09$, $a_3=0.052\pm 0.027$ and $b_1=1.69\pm0.21$. Next, we free $a_4$ imposing that both functions must match at some unknown point $Q_0^2$, i.e., requesting continuity and differentiability at $Q_0^2$ with an error of $10^{-11}$:

\begin{figure}
\includegraphics[scale=0.5]{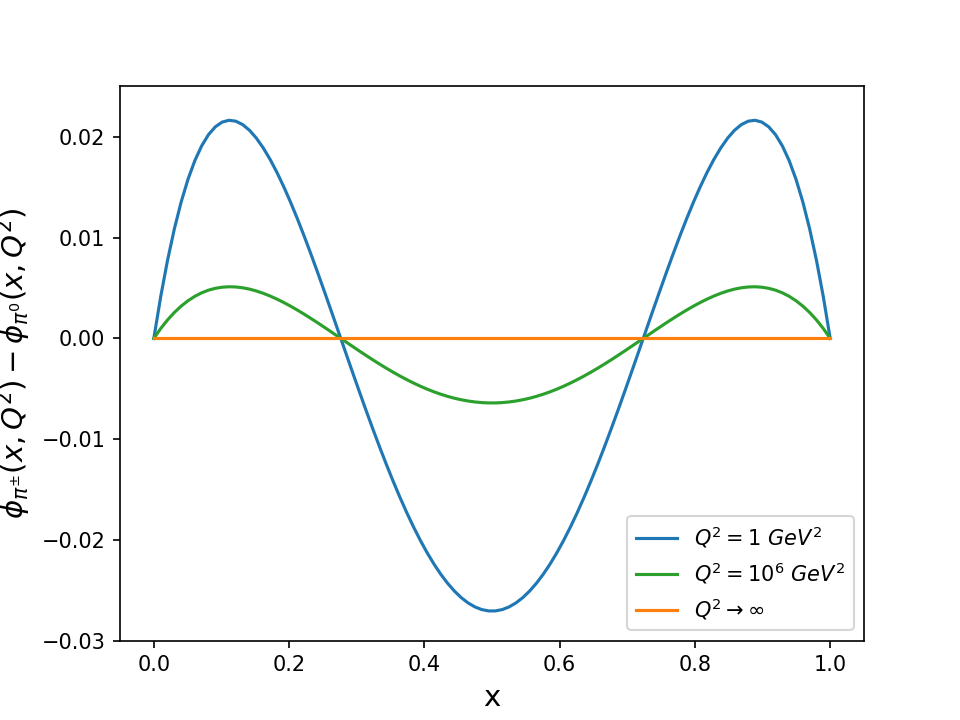}
\caption{Difference between neutral and charged pion DAs, giving the size of isospin breaking effects at $Q^2=1 \text{ GeV}^2$ (blue), at $Q^2=10^6 \text{ GeV}^2$ (green) and the asymptotic limit $Q^2\to\infty$ (orange); this last one is identically zero since neutral and charged pion DAs have the same asymptotic DA.}
\label{fig:IsospinPDA}
\end{figure}

\begin{equation}
P_1^4(Q_0^2, a_4)=Q_0^2F_\pi(Q_0^2)
\end{equation}
\begin{equation}
\frac{d P_1^4(Q^2)}{dQ^2}\Bigg{|}_{Q=Q_0^2, a_4}=\frac{d(Q^2F_\pi(Q^2))}{dQ^2}\Bigg{|}_{Q=Q_0^2}
\end{equation}
Solving the above system numerically, we obtain $Q_0^2=2.21 \text{ GeV}^2$ and $a_4=-0.014$.
In this way, the full function is depicted in Fig.(\ref{fig:chargedFF}), which fits the data with $\tilde{\chi}^2=1.00$ and 115 DOF. As can be seen in the figure, data are described by a curve with a maximum at low energies and then decreasing towards the asymptotic limit Eq.(\ref{eqn:chargedAsympLim}). In a black discontinuous line, the pQCD prediction is shown with an arbitrary height, to show its trend at high energies.  
As in the case of the $F_0(Q^2)$, a large set of PAs have been studied. The $\tilde{\chi}^2$ and AIC selected $P_1^4(Q^2)$ as the best candidate.

Returning to Fig.(\ref{fig:chargedFF}), our fit remains above this pQCD prediction, a feature that was also obtained previously in other numerical approaches \cite{Masjuan}, \cite{ChangCloet}. Finally, the $\phi_{\pi}$ corresponding to $F_{\pi}$ is depicted in Fig.(\ref{fig:PDA}), properly normalized following Eq.(\ref{normalizedDAs}). The results for this DA nicely compare with Ref.\cite{CloetChang}.

\subsection{Isospin-breaking effects from the Pion DAs}

\begin{figure}
\includegraphics[scale=0.5]{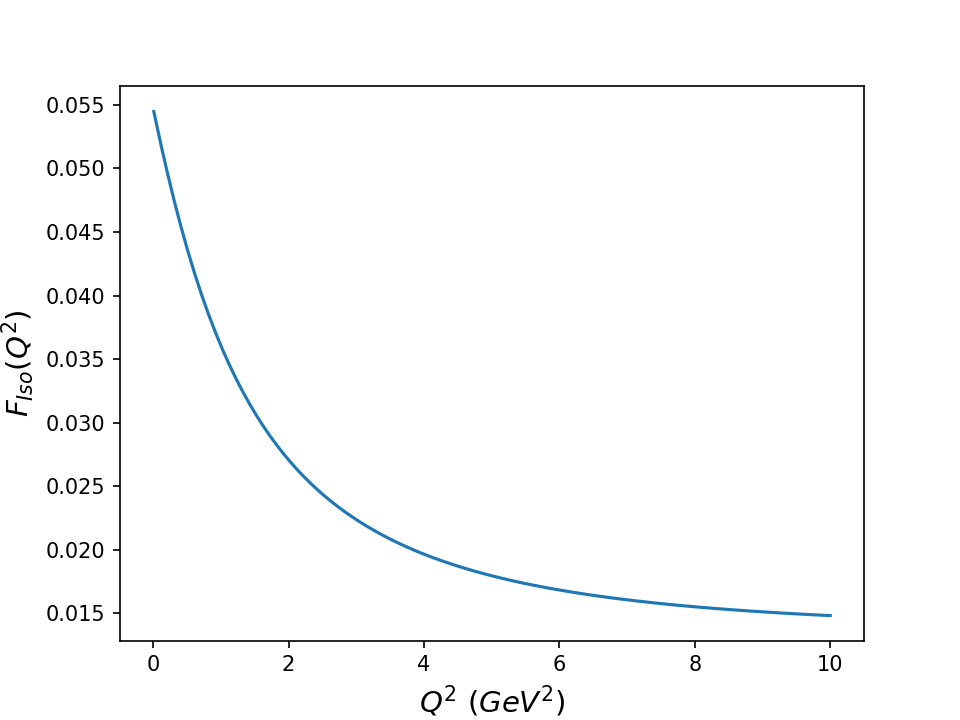}
\caption{Isospin breaking form factor $F_{\rm Iso}(Q^2)$ from Eq.(\ref{eqn:isospinFF}).}
\label{fig:IsospinFF}
\end{figure}

While no direct relation between the pion DAs and the pion mass exist, the difference $\phi_{\pi^{\pm}} - \phi_{\pi^0}$ is directly related to isospin-breaking effects and thus to the quadratic pion mass difference $\Delta m_{\pi}^2 = m_{\pi^{\pm}}^2 - m_{\pi^0}^2$. In this section, we explore this idea using a simple sum rule and the results of the previous subsections. 
 Fig.(\ref{fig:IsospinPDA}) plots the difference of DAs $\phi_{\pi^\pm}(x,Q^2)-\phi_{\pi^0}(x,Q^2)$ at various $Q^2$ values as a function of the momentum fraction $x$. By considering this difference as a distribution amplitude, we can compute its integral, interpreted as an \textit{effective} isospin breaking form factor $F_{\text{Iso}}$, as:

\begin{equation}
\label{eqn:isospinFF}
F_{\text{Iso}}(Q^2)= \int_0^1 \frac{dx}{x}\left(\phi_{\pi^\pm}(x,Q^2)-\phi_{\pi^0}(x,Q^2)\right).
\end{equation}
$F_{\text{Iso}}(Q^2)$ is depicted in Fig.(\ref{fig:IsospinFF}).
In the time-like region the form factor $F_{\text{Iso}}(-Q^2)$ is given by the dispersive integral:
\begin{equation}
\label{eqn:dispersionIntegral}
F_{\text{Iso}}(-Q^2)=\frac{1}{\pi}\int_{s_{th}}^\infty \frac{\text{Im }F_{\text{Iso}}(s)}{s+Q^2+i\epsilon}ds.
\end{equation}

$\text{Im }F_{\text{Iso}}(s)$ is parametrized by the difference of the densities $\rho_{\pi^0}$ and $\rho_{\pi^\pm}$, after isolating $s=-m_{\pi^0}^2$ and $s=-m_{\pi^\pm}^2$ from the continuum term $\rho^{\rm cont}$:
\begin{equation}
\begin{split}
\frac{1}{\pi}\text{Im }F_{\text{Iso}}(s)=f_{\pi^0}^2\delta(s+m_{\pi^0}^2)+\rho_{\pi^0}^{\rm cont}\theta(s-s_{th})\\-f_{\pi^\pm}^2\delta(s+m_{\pi^\pm}^2)-\rho_{\pi^\pm}^{\rm cont}\theta(s-s_{th}).
\end{split}
\end{equation}
$f_{\pi^0}$ and $f_{\pi^\pm}$ are the neutral and charged pion decay constants, respectively. We assume the isospin-breaking effects are negligible for the continuum part such that they cancel out, so we consider:
\begin{equation}
\frac{1}{\pi}\text{Im }F_{\text{Iso}}(s)\approx f_{\pi^0}^2\delta(s+m_{\pi^0}^2)-f_{\pi^\pm}^2\delta(s+m_{\pi^\pm}^2).
\end{equation}
Using this expression in Eq.(\ref{eqn:dispersionIntegral}) we obtain:
\begin{equation}
\label{eqn:twoPoles}
F_{\text{Iso}}(-Q^2)=\frac{f_{\pi^0}^2(Q^2+m_{\pi^\pm}^2)-f_{\pi^\pm}^2(Q^2+m_{\pi^0}^2)}{(Q^2+m_{\pi^0}^2)(Q^2+m_{\pi^\pm}^2)}.
\end{equation}
By evaluating this function at $Q^2=0$ we can relate $F_{\text{Iso}}(0)$ with $\Delta m_{\pi}^2=m_{\pi^\pm}^2-m_{\pi^0}^2$. The value $F_{\text{Iso}}(0)=0.055$ is computed from Eq.(\ref{eqn:isospinFF}). A naive estimate of the uncertainty takes into account the uncertainty associated to both pion DAs from the corresponding fits. Adding them up linearly, a rough estimate of about $10\%$ is found. Assuming $f_{\pi^0}^2\approx f_{\pi^\pm}^2$ in Eq.(\ref{eqn:twoPoles}), we obtain the approximate relation:
\begin{equation}
\Delta m_{\pi}^2=F_{\text{Iso}}(0)\frac{m_{\pi^0}^2m_{\pi^\pm}^2}{f_{\pi}^2},
\end{equation}
which numerically, using the inputs $F_{\text{Iso}}(0)=0.055(5)$ from Eq.(\ref{eqn:isospinFF}) and the PDG values for $m_{\pi^0}$ and $m_{\pi^\pm}$ \cite{Navas}, we obtain $\Delta m_{\pi}^2=(1.1\pm0.1)\cdot 10^{-3} \text{ GeV}^2$, to compare with the experimental result $\Delta m_{\pi}^2|_{\text{exp}}=1.3\cdot 10^{-3} \text{ GeV}^2$.

\section{Conclusions}
In this work, we have explored an application of a model for the strong coupling constant Eq.(\ref{eqn:DDSW}) obtained in \cite{CancioMasjuan}, to the neutral pion transition form factor and to the charged pion form factor. We have seen that in both cases our $\hat{\alpha}_s(Q^2)$ have influence in the intermediate-energy region, meaning that scales traditionally considered as high energies could be influenced by non-perturbative effects. This is consistent with the experimental data of both form factors. Neutral pion transition form factor is puzzling in the sense that experimental data at high energies is above the predicted asymptotic limit from pQCD, but at the same time our expansion does not favour a maximum in this energy region. On the other hand, our results for the charged pion form factor presents such a maximum and then a decreasing well above the pQCD prediction, favouring that we are missing some effects in this region, which can be taken into account by an appropiate resummation of the strong coupling constant. 

A study of the isospin breaking effects induced by the difference of DAs for charged and neutral pions is presented, including a determination of the quadratic pion mass difference $\Delta m_{\pi}^2$ with a reasonable comparison with the current experimental value.

Finally, it is clear that $\Lambda_{QCD}$ is well above than previously thought. The exercise as proposed here opens the door to a full phenomenological field of studies.

\section*{Acknowledgments}
This work has been supported by the Ministerio de Ciencia e Innovación under grant PID2020-112965GB-I00, by the Secretaria d’Universitats i Recerca del Departament d’Empresa i Coneixement de la Generalitat de Catalunya under grant 2021 SGR00649, and by the Spanish Ministry of Science and Innovation (MICINN) through the State Research Agency under the Severo Ochoa Centres of Excellence Programme 2025–2029 (CEX2024-001442-S). IFAE is partially funded by the CERCA program of the Generalitat de Catalunya. 

\nocite{*}
\bibliographystyle{unsrt}

\end{document}